\documentclass[12pt]{iopart}
\usepackage{amsfonts, amsmath, amssymb, amsthm, mathtools}
\usepackage{graphicx}
\usepackage[numbers]{natbib}
\newtheorem{example}{Example}[section]

\begin{document}

\title[DNA Digital Data Storage and Retrieval using algebraic codes]{DNA digital data storage and retrieval using algebraic codes}
\author{NallappaBhavithran G$^1$ and Selvakumar R$^{2, *}$}

\address{1 Vellore Institute of Technology, Vellore, India. Email : nallabhavitran@yahoo.com}

\address{2 Vellore Institute of Technology, Vellore, India. Email : rselvakumar@vit.ac.in}

\vspace{10pt}
\address{* Corresponding author}

\begin{abstract}
DNA is a promising storage medium, but its stability and occurrence of Indel errors pose a significant challenge. The relative occurrence of Guanine(G) and Cytosine(C) in DNA is crucial for its longevity, and reverse complementary base pairs should be avoided to prevent the formation of a secondary structure in DNA strands. We overcome these challenges by selecting appropriate group homomorphisms. For storing and retrieving information in DNA strings we use kernel code and the Varshamov-Tenengolts algorithm. The Varshamov-Tenengolts algorithm corrects single indel errors. Additionally, we  construct codes of any desired length (n) while calculating its reverse complement distance based on the value of n.
\end{abstract}

%
\vspace{2pc}

\section{Introduction} 
\begin{figure}[hbt!]%
    \centering
    \includegraphics[width=0.9\textwidth]{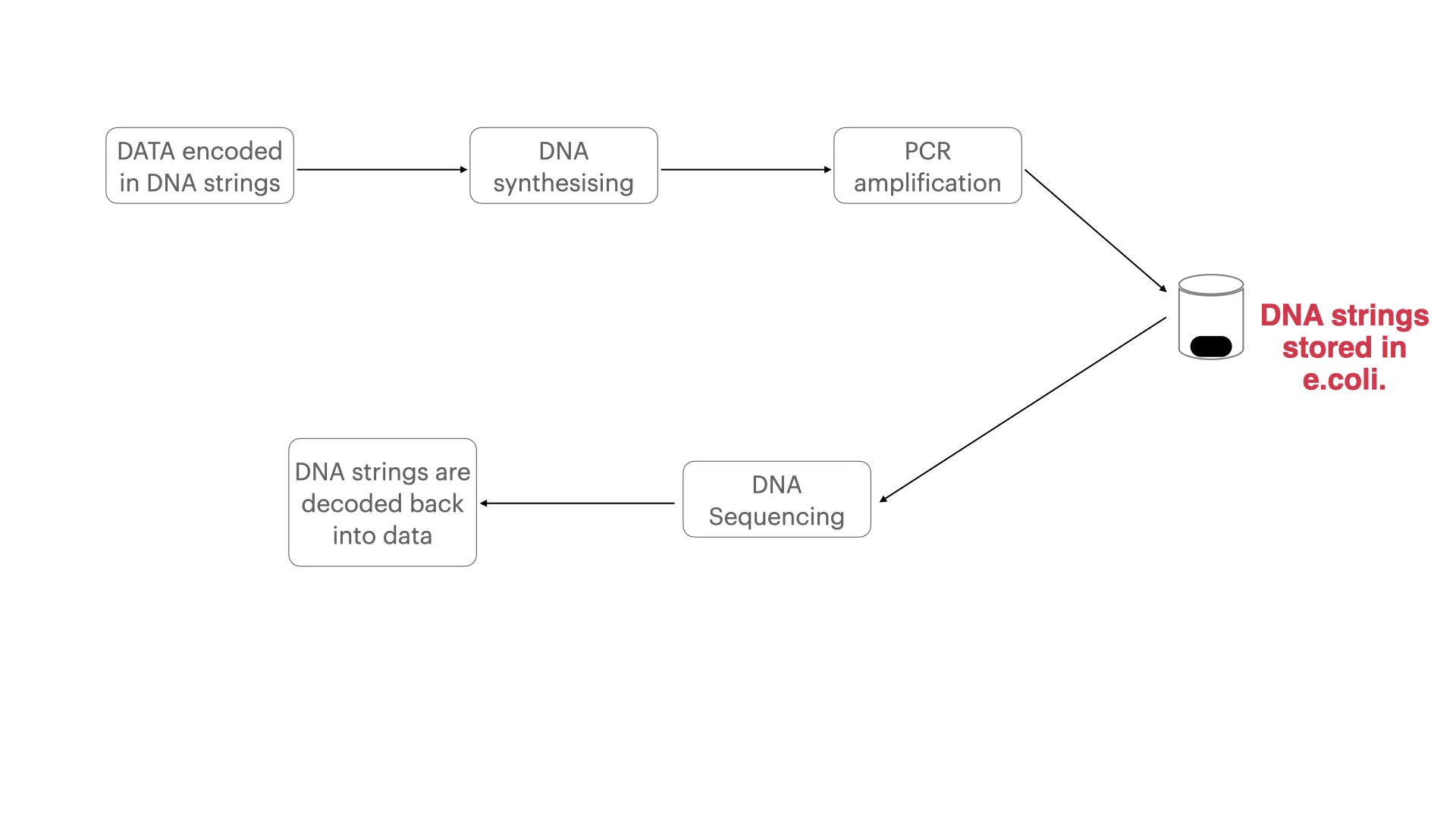}
    \caption{Systematic procedure of DNA-based storage systems}
    \label{fig:systematic_procedure}
    \noindent
  \end{figure}

Over the last few years, there has been significant growth in the creation and replication of data, and this tendency will continue. Currently, large-scale data is stored using silica-based storage systems, which come with high maintenance costs. One of the reasons behind the high maintenance costs of storage is that silica-based storage devices have a lifespan of approximately thirty years. Thus, they need to be updated every ten to twenty years. Furthermore, the disposal of outdated systems is complex due to silicon's non-biodegradable nature. To address the challenge of storing exponentially growing data, it is vital to find a storage technology that is both convenient and affordable. DNA-based storage technology presents an alternative solution. The idea of storing information in DNA was first proposed by C.T. Clelland in 1999\cite{bi2}. The human genome can be used as a way to hide information since it is challenging to identify the gene to crack. Additionally, an elliptic curve encryption algorithm was proposed to encrypt information and hide it on the DNA sequence\cite{bi28}. Similarly, a method that employs Huffman encoding and XOR is used to encrypt information. Moreover, the discovery that information can be recovered from sources dating back thousands of years led to the development of DNA-based storage systems\cite{bi29}. Compared to silica-based storage systems, DNA-based storage systems are more durable and capable of handling exponentially growing data. It is possible to store exabytes of data in a single gram of DNA. Although the cost of synthesizing and sequencing DNA storage systems is substantial, this cost is expected to decrease in the coming years. In recent years, the amount of data created and replicated has significantly increased, and this trend is projected to continue. Currently, silica-based storage systems are used to store large-scale data, but they come with high maintenance costs due to their limited lifespan. Silica-based devices typically need to be updated every ten to twenty years, and their disposal is complicated because silicon is non-biodegradable. To overcome the challenge of storing exponentially growing data, it is essential to find a storage technology that is both convenient and affordable. DNA-based storage technology provides a promising alternative solution. The concept of storing information in DNA was first proposed in 1999, and it has since been developed into a viable option. The human genome can be used to hide information because it is difficult to identify the gene to crack. Various encryption algorithms, such as elliptic curve and Huffman encoding, can be used to encrypt information and hide it in DNA sequences. Additionally, the discovery that information can be retrieved from sources dating back thousands of years has led to the creation of DNA-based storage systems. Compared to silica-based storage systems, DNA-based systems are more durable and capable of handling vast amounts of data. A single gram of DNA can store exabytes of data. Although synthesizing and sequencing DNA storage systems are currently expensive, the cost is expected to decline in the future.\cite{bi1}\\
The structure of DNA is a double helix, which is made up of two connected strands that loop around one another to resemble a twisted ladder. The backbone of each strand is composed of alternating deoxyribose and phosphate groups. Adenine (A), cytosine (C), guanine (G), or thymine are the four bases that are joined to each sugar (T). Those two strands of nucleotides are held together by hydrogen bonds between the complementary bases A = T (double bond) and C $\equiv$ G(triple Bonds) and denoted as A$^c$ = T, G$^c$ = C and vice versa. Information is encoded into A, G, T, and C sequences, and the corresponding nucleotides are synthesised(oligos). These oligos are stored in e.coli and can be retrieved later. Figure \ref{fig:systematic_procedure} illustrates the entire process of storing and retrieval of data. The oligos typically include 100 to 1000 nucleotide bases since synthesising them is expensive and prone to mistakes. Polymerase Chain Reaction(PCR) is used to amplify the stored information and create numerous copies of the same information. A DNA sequencer aligns these sequences with retrieving the data.\cite{bi5} \cite{bi27}\\

Starting from kilobytes in 2012 \cite{bi6} \cite{bi7} \cite{bi30}researchers created coding schemes that managed to store and retrieve megabytes \cite{bi8} \cite{bi9} of information from DNA. In 2016, Yazdi \cite{bi10} devised a system to store six universities’ budgets that could be randomly accessed and modified. To get random access, they placed addresses on both ends of 27 DNA strands, which were used to store the university information. These addresses contain 20 nucleotide bases and are self-uncorrelated to each other. Additionally, each piece of information is encoded into three nucleotide bases(A, T and C) dependent on their address, and the modification is done using OE-PCR and g-block technologies.\\
To rectify errors made during DNA synthesis and sequencing, a suitable error-correcting code \cite{bi3} must be devised. The error-correcting codes are created by adding redundancy bits to the information while encoding the information. These redundancy bits are helpful in retrieving information without errors. The set of all encoded information is called the code, and each element of the code is called a codeword. The encoding is done in such a way as to preserve some of the algebraic characteristics of the information. One of the primary focuses of error in DNA systems is indel errors. Many bounds have been achieved for the indel errors with the usage of error correction codes.\\
Levenshtein\cite{bi4} suggested a method for recovering the original data even after the codeword had a single bit removed. After half a century Brakensiek \cite{bi24} generalised the result for multiple deletions. However, for more than 2 deletions the code is hard to design.  Sima \cite{bi25} and Song \cite{bi26} also suggested different coding schemes and attained an optimal bound for retrieving k-deleted bits from the codeword.\\
This article presents a new technique for eliminating Indel errors in DNA strings. Additionally, this technique imposes GC-content and reverse complement constraints on the code to improve DNA string stability and prevent secondary structure formation. In section \ref{sec2}, we discuss the problems which arise while using DNA as a storage system.  In section \ref{sec3}, we analyse those problems discussed in section \ref{sec2} by imposing constraints. Finally, in section \ref{sec5}, we gave a procedure by using VT and kernel code to construct an indel error-free DNA code.
\section{DNA storage model}\label{sec2}
A DNA code $\mathcal{C}_{DNA}(n, M) \subset \{A, G, T, C\}^n$ (nucleotide bases) with each DNA codeword of length n and size M. Let $x, y \in \mathcal{C}_{DNA}(n, M)$ The Hamming distance H($x, y$) between two codewords is the number of distinct elements in those two codewords. 
The \textit{reverse} of a codeword $x = (x_1, x_2, \dots, x_n)$ is $(x_n, x_{n-1}, \dots, x_2, x_1)$ and denoted by $x^R$. Similarly the \textit{reverse compliment}  of a codeword $x = (x_1, x_2, \dots, x_n)$ is $(x_n^c, x_{n-1}^c, \dots, x_2^c, x_1^c)$ and denoted by $x^{RC}$.
For example: Let us consider n = 3, x = (A, G, C), y = (A, T, G) then H(x, y) = 2 since second and third position of x and y are different, $x^R$ = (C, G, A) and $x^{RC}$ = (C$^c$, G$^c$, A$^c$) = (G, C, T).
Whenever a storage system is considered, the occurrence of errors is unavoidable. In DNA-based storage systems, errors occurring during synthesis, sequencing, and PCR amplification are referred to as indel, sequencing, and storing errors, respectively.
\subsection{Indel Errors}
The deletion, insertion, and substitution errors that take place during DNA synthesis are referred to as indel errors. In the deletion, a nucleotide base is deleted from the original DNA string, and the position of deletion is random. In the insertion, a nucleotide base is inserted into a random place in the original DNA string. The insertion and deletion change the length of the original sequence. Substitution errors randomly replace some of the bases in the original string.
For example: Let x = ATACGTTCA, and a base G is inserted into the 2nd position. The resulting word would be A\textbf{G}TACGTTCA. If the 4th base is deleted, the resulting string will be AT\textbf{AG}TTCA. If a substitution of A to G is in third place will be AT\textbf{G}CGTTCA.
Levenshtein distance$(d_l(x, y ))$ is introduced to find the relation between two strings with respect to indel errors. For example, let x = ATACGTTCA, y = AGTAGTTCA, then $d_l(x, y)$ = 2, since deleting the second position ‘G’ and inserting a ‘C’ in the fifth position converts y to x. However, Hamming distance is used to determine the relationship between two DNA sequences since calculating the Levenshtein distance is NP-hard.
\subsection{Sequencing error}
In sequencing, small fragments are pieced together to create the whole sequence. Tatum error is a sequencing error that occurs when a single base pair or section of the DNA string is replicated and placed adjacent to the original. As an example, suppose ATACGTCA is a DNA string. Upon sequencing, one would receive ATATACGTCA. The length two tantum error occurred as the first two base pairs “AT”. Most of the time, this error correction is done for therapy and other medical purposes. But we also need to consider this when talking about the longevity of DNA and other materials that need to be stored for an extended period.\cite{bi16}
\subsection{Storing Errors}
For replicating the DNA, a process called hybridisation is used. However, the unwanted hybridisation in long strands results in the formation of a secondary structure. Since DNA strings are sequenced to form the long strand of a DNA molecule, complementary substrings of the concatenation of two DNA strings should not be in the long strand. (i.e.) Let x = AGT, y = CAG, then any substring of (xy)$^{RC}$ = CTGACT should not be a codeword of $\mathcal{C}_{DNA}(n, M)$. For the stability of DNA proportionate amount of G and C has to be present in the codeword. The number of occurrences of G or C in a codeword(say x) is called the GC weight of the codeword and is denoted by $w_{GC}(x)$. The GC weight ratio to the codeword's length is called the codeword's GC content.
\section{Constraints on DNA codes}\label{sec3}
\subsection{Hamming Distance constraint}
Let $\mathcal{C}$ be a \textit{DNA code}, $H(x, y) \geq d,$ $ \forall x,y \in \mathcal{C}$ with $x\neq y$. We refer to this distance as the minimum distance for Code $\mathcal{C}$. This acts as an alternative to the Levenshtein distance.
\subsection{Reverse constraint}\label{Reversecon}
 Let $\mathcal{C}$ be a \textit{DNA code}, $H(x^R, y) \geq d,$ $ \forall x,y \in \mathcal{C}$.Also considering the case $x = y$. This step acts as a bridge for constructing the reverse complement constraint. One can use the bound of this constraint and implement it on the other. 
 \subsection{Reverse Complement constraint}\label{Reversecomple}
 Let $\mathcal{C}$ be a \textit{DNA code}, $H(x^{RC}, y) \geq d,$ $\forall x,y \in \mathcal{C}$.Also considering the case $x = y$. This constraint helps in the reduction of unwanted hybridization errors.
 \subsection{GC-content constraint}\label{GCcon}
 PCR amplification is difficult if the GC content is high, whereas a low GC content results in a non-stable gene. So, the GC content of all the codewords in a DNA code C should be a fixed number, say w. \\
 These constraints are imposed in the DNA code to eradicate the errors discussed in section 2. To impose these constraints, we are using VT and concatenated kernel code \cite{bi17} \cite{bi18} to satisfy these constraints.
 \section{Construction of DNA codes}\label{sec5}
\begin{figure}[hbt!]%
    \centering
    \includegraphics[width=0.9\textwidth]{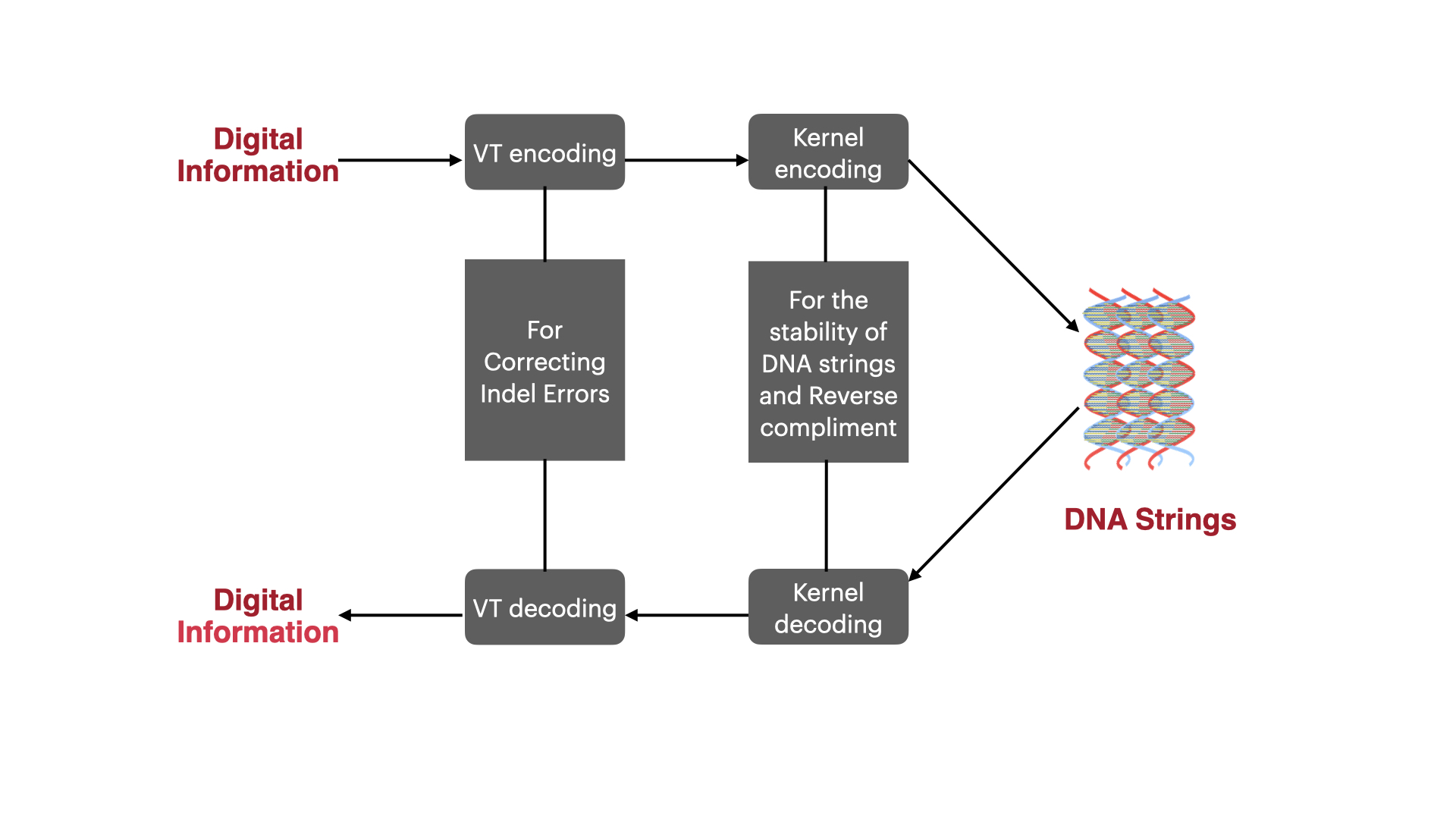}
    \caption{Our encoding and decoding procedure}
    \label{fig:our_procedure}
  \end{figure}

  Here, a code of length n is constructed with a message block of length: \begin{equation}
      l = n - log_2(2 * n - 1) - 1 \label{eq:messagebit}
  \end{equation}
  Each element of the block is from the same finite abelian group $\mathcal{G}$. To construct an indel error-free DNA code with GC-constraint(section \ref{GCcon}) and Reverse Complement constraint(section \ref{Reversecomple}) VT code and kernel code have been used to encode the information (figure\ref{fig:our_procedure}). The algorithm goes as follows:
  \begin{enumerate}
      \item The set of all information of length $l$ is generated.
      \item Each of that information is encoded using VT code to restrict indel errors.
      \item Each VT-encoded information is then mapped to a unique element in the Kernel code of length n+1.
      \item Now, the mapped elements of the sets are again outer encoded using the proposed homomorphism to address the constraints (section \ref{sec3}).
      \item Then, the obtained word is mapped onto the corresponding base pairs.
  \end{enumerate}
  The construction begins by encoding the information using VT codes. VT codes are used for correcting single deletion, insertion or substitution(indel) errors. Goldman \cite{bi14} has suggested a way for correcting multiple indel errors by segmenting codewords into smaller lengths.\\
 The first step of the algorithm is calculating the maximum length for the message bits($l$) equation (\ref{eq:messagebit}). All possible information on length $l$ is encoded using VT code. Then, the set of all VT encoded information(say Y) is obtained. 
 \begin{example}\label{1stexm}
  Example: For $n=10$, the value $l$ is 4. Since, $n-1=9$ is not a power of 2, the par\_pos (parity position) value is [$2^0, 2^1, 2^2, 2^3, 9$].\\
  Let our information $\mathbf{a}$ be $1011$, then in the VT encoding $\mathbf{a}$ first mapped onto the non-parity positions  $001001100$(say $y$). The syndrome of $y$ (i.e. 3) is partitioned as 2+1. So, the VT-encoded information of $a$ is $111001100$ (say $y_{VT(a)}$).
  \end{example}
  In the second step, the kernel encoding is performed with a length of $n+1$ for the VT-encoded information. Prior to encoding the information, the definition of kernel code is given below.\\
   \textbf{Definition:} Let $\mathcal{G} = \mathcal{G}_1 \times \mathcal{G}_2 \times \cdots \times \mathcal{G}_n$, where $\mathcal{G}_i's$ are groups and ($\mathcal{S}, *$) be an abelian group with identity element \textit{e}. Then $\mu : \mathcal{G} \rightarrow \mathcal{S}$ such that $\mu(g_1, g_2, \dots, g_k) = \mu_1(g_1)*\mu_2(g_2)* \dots *\mu_n(g_n)$, where $g_i \in \mathcal{G}_i$ and $ \mu_i: \mathcal{G}_i \rightarrow S$ is a homomorphism for all $ i = 1\text{ }to\text{ }n$. This construction states that $\mu$ is a homomorphism. The kernel of the homomorphism $K = \{g \in \mathcal{G}/\mu(g) = e\} $ is called the Kernel Code(\cite{bi17}).\\
   For our encoding, we consider all $\mathcal{G}_i$ and $S$ to be $\mathcal{Z}_2$. We only take into consideration the subset($S$) of the kernel of the homomorphism that starts with one.
   \begin{example}\label{2ndexample}
       The VT-encoded information $y_{VT(a)} = 111001100$ from example \ref{1stexm} is mapped to $s_{k(y)} = 11110011000$. Clearly $s_k(y)$ is an element of $S$.
   \end{example}
    In the third step, n - 1 redundancy bits are added to each element $s_{k(y)}$ of the kernel subset S and denoted as $E_y$. Here, $i^{th}$ redundancy bit is obtained using the homomorphism $h_i$. The homomorphism goes as follows.
    \begin{equation}
        h_i(g_1, g_2, \dots, g_n) = \begin{cases} 
        g_{i+1} & 1 \leq i \leq \lfloor{\frac{n-1}{2}}\rfloor\\
        g_1 + g_{i+1} & \lceil{\frac{n+1}{2}}\rceil \leq i\leq n-1 \\
        {g_1 + g_{i+1} + g_{n+1}} & i = \frac{n}{2}, \\
        &\frac{n}{2} \text{ is an integer} 
    \end{cases}
    \end{equation}
    \begin{example}\label{3rdexm}
        From the example (\ref{2ndexample}) $s_{k(y)} = 11110011000$ the concatenated Kernel encoding is $E_y = 11110011000111000011$.
    \end{example}
    In the final step, the binary codeword is separated into small strings of length two where the bit $i$ is paired with $(i + n)^{th}$ where i ranging from 1 to n and mapped on to one of the four base pairs. Here, the mapping are as follows: $'00' \rightarrow C,\text{ } '01' \rightarrow A,\text{ } '10' \rightarrow T, \text{ }'11' \rightarrow G$. This binary conversion is dependent on an element from each half. In this way, $l$ length information is mapped onto the n-length bps.\\ 
    \begin{example}\label{4thexm}
        For $E_y = 11110011000111000011$(example \ref{3rdexm}), the DNA encoded codeword is $TGGGCCTTAA$. 
    \end{example}
    
    \begin{figure}[ht!]%
        \centering
        \includegraphics[width=0.9\textwidth]{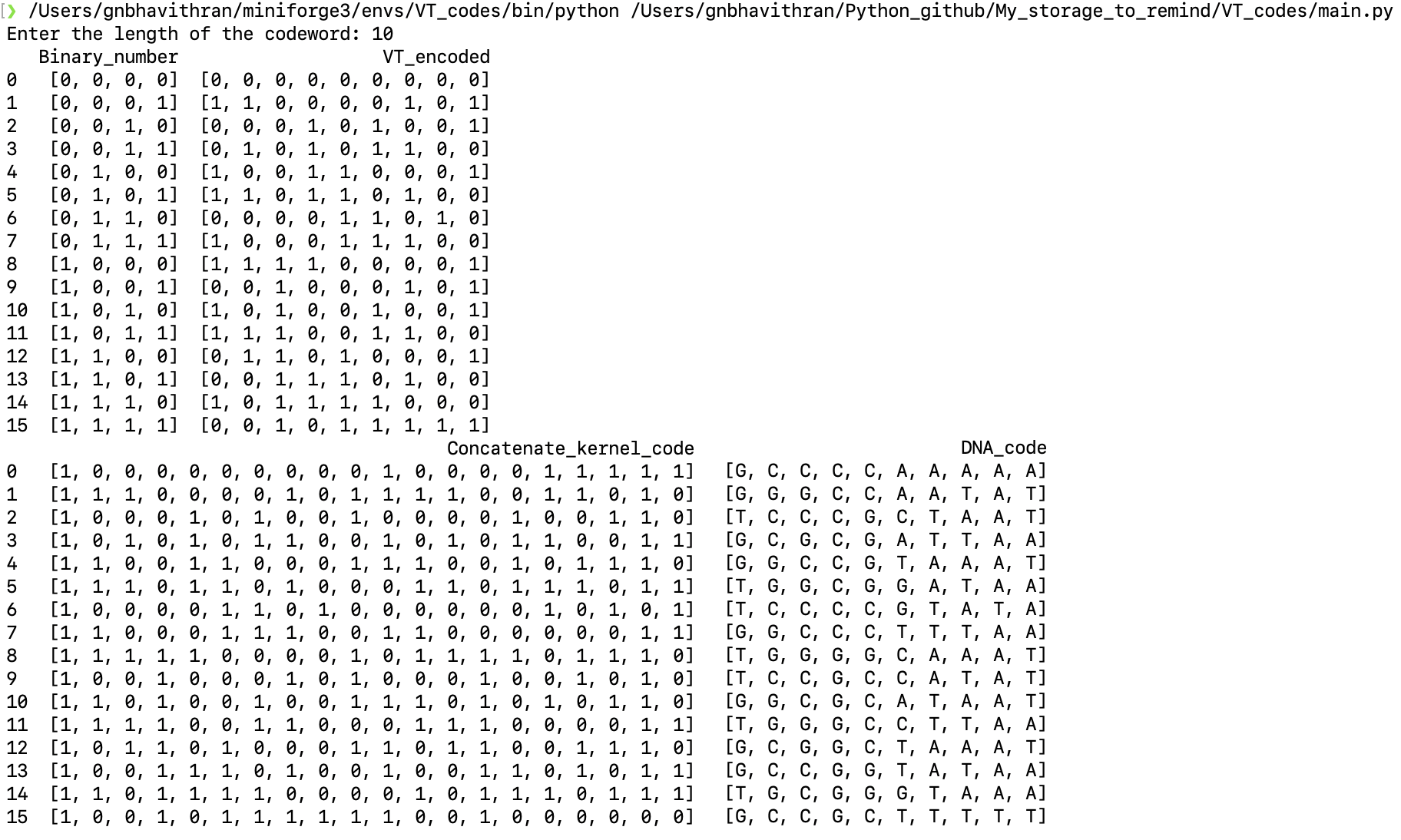}
        \caption{Encoding for three-length information set}
        \label{fig:my_label}
        \noindent\footnotesize{Source: It is the python code output  which was executed on VSC (open source software) on the computer with the processor mac m1}
    \end{figure}

    When retrieving information, the second-to-last base pairs are mapped to $\mathcal{Z}_2$. 'A' and 'T' are mapped to '0', and 'G' and 'C' are mapped to '1'.
    \begin{example}
        Suppose the DNA codeword from the example \ref{4thexm}($TGGGCCTTAA$) is received as $r = TGGCCTTAA$. Then, the $\hat{r} = 11001100$. The syndrome and the weight of the $\hat{r}$ are 5 and 4 respectively. Since weight is less than syndrome $1$ has to be added before the $(5-4)^{th}$ zero. So, the decoded value will be $\hat{r_{VT}} = 111001100$. From that, information is only in the non-parity bits(from example \ref{1stexm}). So, $\hat{a} = 1011$.
    \end{example}
    Figure \ref{fig:my_label} depicts how the encoding has been done for an information set of length(l) 4 and codeword(n) of length 10 using Python3.\\
    This way of encoding preserves the GC-content of $50\%$ for even length n and odd length, it ranges from $40 \% $ to $ 60\%$. This determines the stability of the DNA code. 
    This also gives a greater distance for reverse complement constraint. For an n-length DNA code, the  RC-distance is $$d_{RC} = 2 \times \Bigg\lfloor{\frac{n-3}{2}}\Bigg\rfloor$$ This acts as an improvised coding scheme for \cite{bi14} where the $3^l$ and other $S_l$ methods have been used. This meets the constraints described in \ref{sec3} and serves as a solution to the DNA storage problem.\\

     \begin{table}[hbt!]
    \centering
    \begin{tabular}{|p{0.2\linewidth} | p{0.3\linewidth}|p{0.15\linewidth} | p{0.15\linewidth}|}
    \hline
         & GC-content & RC length constrain & Indel Error Correction  \\
         \hline
         Linear and nonlinear constructions of DNA codes with Hamming distance d, constant GC-content and a reverse-complement constraint (\cite{bi13})& constant & Yes & No\\
         \hline
         On multiple-deletion multiple-substitution correcting codes. (\cite{bi11}) & No & No & Yes\\
         \hline
         Rewritable random access (\cite{bi10})& Maintained only for address with even length & No & No\\
         \hline
         Our paper & Maintained & Yes & Yes\\
         \hline
    \end{tabular}
    \caption{Effectiveness of Kernel encoding}
    \label{tab:comparision_others}
\end{table}

The comparison presented in Table \ref{tab:comparision_others} shows that our approach is more efficient in generating code that adheres to RC constraints at any distance (as shown in equation \ref{eq:messagebit}). Our code is capable of maintaining the GC content between 40\% and 60\% for any length and also addresses single insertion or deletion errors.

    \section{Conclusion}\label{sec13}
    As DNA has a high chance of occurring for mutation errors and instability, algorithms for generating balanced GC content code have been proposed. The balanced GC- content ensures the stability of DNA strings. Further, the VT algorithm used here corrects a single indel error that occurs in DNA strings during retrieval.
    
    \bibliographystyle{unsrt}
    \bibliography{bhavithran_DNA_Paper}
\end{document}